\documentclass[12pt]{elsarticle}

\makeatletter

\usepackage{graphicx}
\usepackage{amsmath}
\usepackage{physics}
\usepackage{url}
\usepackage{multirow}
\usepackage{siunitx}
\usepackage{graphbox}
\usepackage{subcaption}
\usepackage{bm}
\usepackage{bbm}
\usepackage{url}
\usepackage{hyperref}

\providecommand{\doi}[1]{%
  \begingroup
    \let\bibinfo\@secondoftwo
    \urlstyle{rm}%
    \href{http://dx.doi.org/#1}{%
      \discretionary{}{}{}%
      \nolinkurl{#1}%
    }%
  \endgroup
}

\newcommand{\rawmatrix}{$(E_x, E_{\gamma})$}
\newcommand{\unfoldedmatrix}{$(E'_x, E'_{\gamma})$}

\DeclareMathOperator*{\argmin}{arg\,min}

\newcommand{\pixtopix}{\texttt{Pix2Pix}}

\newcommand{\cade}[1]{{{\textcolor{black}{#1}}}}

\usepackage{hyperref}
\graphicspath{ {images/} }
\usepackage{amssymb}

\journal{Nuclear Instruments and Methods A}

\begin{document}

\begin{frontmatter}


\title{Two-dimensional total absorption spectroscopy with conditional generative adversarial networks}



\author[MSUPA,FRIB,JINA]{C. Dembski}
\author[DavPhy,DavCS]{M.P. Kuchera}
\author[FRIB,MSUCEM]{S. Liddick}
\author[DavCS]{R. Ramanujan}
\author[MSUPA,FRIB,JINA]{A. Spyrou}
\address[MSUPA]{Department of Physics and Astronomy, Michigan State University, East Lansing, MI, 48824, USA}
\address[FRIB]{Facility for Rare Isotope Beams, Michigan State University, East Lansing, MI, 48824, USA}
\address[JINA]{Joint Institute for Nuclear Astrophysics, Michigan State University, East Lansing, MI, 48824, USA}
\address[DavPhy]{Department of Physics, Davidson College, Davidson, NC, 28035, USA}
\address[MSUCEM]{Department of Chemistry, Michigan State University, East Lansing, MI, 44824, USA}
\address[DavCS]{Department of Mathematics and Computer Science, Davidson College, Davidson, NC, 28035, USA}

\begin{abstract}

We explore the use of machine learning techniques to remove the response of large volume $\gamma$-ray detectors from experimental spectra. Segmented $\gamma$-ray total absorption spectrometers (TAS) allow for the simultaneous measurement of individual $\gamma$-ray energy (E$_\gamma$) and total excitation energy (E$_x$). Analysis of TAS detector data is complicated by the fact that the E$_x$ and E$_\gamma$ quantities are correlated, and therefore, techniques that simply unfold using E$_x$ and E$_\gamma$ response functions independently are not as accurate. In this work, we investigate the use of conditional generative adversarial networks (cGANs) to simultaneously unfold $E_{x}$ and $E_{\gamma}$ data in TAS detectors. Specifically, we employ a \texttt{Pix2Pix} cGAN, a generative modeling technique based on recent advances in deep learning, to treat \rawmatrix~ 
matrix unfolding as an image-to-image translation problem. We present results for simulated and experimental matrices of single-$\gamma$ and double-$\gamma$ decay cascades. Our model demonstrates characterization capabilities within detector resolution limits for upwards of 93\% of simulated test cases.
\end{abstract}


\end{frontmatter}


\section{Introduction}
\label{Intro}

Peak finding is a primary step in many forms of spectroscopic analysis and is used in a number of domains such as molecular identification \cite{Elyashberg2015, Yang2018}, the study of distant, high-redshift galaxies \cite{Lowenthal1997,Lehnert2010}, and in applications across nuclear sciences \cite{Crouthamel2013}.
Peak finding is a problem well-suited to automated analysis methods and the ability of modern deep learning networks to efficiently analyze one-dimensional spectra has been shown in $\gamma$-spectroscopy \cite{Jhung2020,Galib2021,WU2021165262} as well as in similar applications like NMR spectroscopy \cite{Cobas2020}. However, these artificial peak-isolation techniques have yet to be applied to two-dimensional spectroscopy data, an important extension necessary to account for correlations between simultaneously measured parameters. An example of such correlated parameters appears in the technique of $\gamma$-ray total absorption spectroscopy (TAS) \cite{Rubio2009, Fujita2011}. TAS measurements can provide both the individual $\gamma$-ray energy ($E_\gamma$) and total excitation energy ($E_x$), two parameters that are not independent from each other.  

Total absorption spectroscopy is a technique used to measure all $\gamma$ transitions associated with the de-excitation of excited states in a nucleus populated in $\beta$-decay. The detection of entire decay cascades, as opposed to only individual $\gamma$-rays as in traditional high-resolution spectroscopy, makes total absorption spectroscopy methods significantly less susceptible to error resulting from the Pandemonium effect \cite{Hardy1977} and makes them particularly well-suited for the measurement of $\beta$-intensity distributions \cite{Rubio2009, Fujita2011}. Multiple TAS detectors are active in current low-energy nuclear physics research, including the Decay Total Absorption Spectrometer (DTAS) detector \cite{Tain2015}, Modular Total Absorption Spectrometer (MTAS) detector \cite{Karny2016}, and Summing NaI(Ti) (SuN) detector \cite{Simon2013}. Total absorption spectroscopy is being used to study $\beta$-strength distributions for applications in nuclear structure \cite{Rasco2016,Rasco2017,Guadilla2016}, reactor decay heat \cite{Fijalkowska2017,Algora2010,Guadilla2017}, and nuclear parameters relevant to astrophysical applications via techniques like the $\beta$-Oslo method \cite{Spyrou2014, Spyrou2017, Liddick2016, Liddick2019}.  TAS detectors have also been used to measure capture reaction cross sections for astrophysical calculations \cite{Spryou2007, Reingold2019}. This work is based on measurements and simulations using the SuN detector at Michigan State University \cite{Simon2013}. SuN is a NaI detector, 16 inches in height and 16 inches in diameter, with a $1.8$ inch wide bore hole along its central axis. Additionally, SuN is segmented into eight optically isolated segments, which provide a measure of the individual $\gamma$ rays participating in a cascade, while summing the total energy deposited in the detector is sensitive to the excitation energy.

A common method of consolidating the multivariate data collected by TAS detectors is to treat the $\gamma$ energy and nuclear excitation energy spectra as $x$- and $y$-axis projections, respectively, of a two-dimensional \rawmatrix~matrix. Between the two axes, these matrices contain crucial information about the entirety of a nucleus's level scheme and are indispensable to total absorption spectroscopy. 

\begin{figure}[ht]
    \centering
    \includegraphics[width=\textwidth]{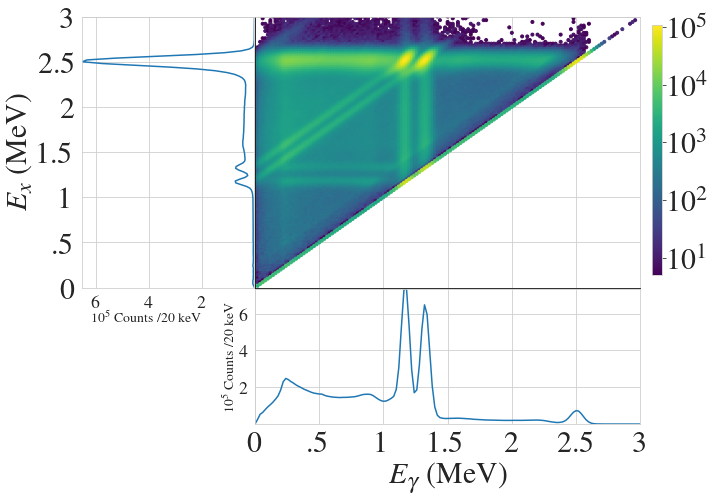}
    \caption{An \rawmatrix~matrix for the decay of $^{60}$Co with $x$ and $y$ axis 1D projections, as measured by SuN.}
    \label{fig:spectra}
\end{figure}

For example, Fig.~\ref{fig:spectra} shows such a 2D matrix that was created from the measurement of a $^{60}$Co radioactive source placed at the center of the SuN detector. The 2D matrix has the excitation energy on the $y$-axis and the individual segment energy on the $x$-axis. The projections of the two axes are also presented in the figure. $^{60}$Co decays predominantly into a level of $^{60}$Ni located at $2505$ keV excitation energy, which is visible in the $y$-axis projection of Fig.~\Ref{fig:spectra}. This level de-excites via the emission of two sequential $\gamma$ rays with energies of $1173$ and $1332$ keV, which are visible in the $x$-axis projection. 

\begin{figure}[htp]

\centering
\includegraphics[width=.3\textwidth]{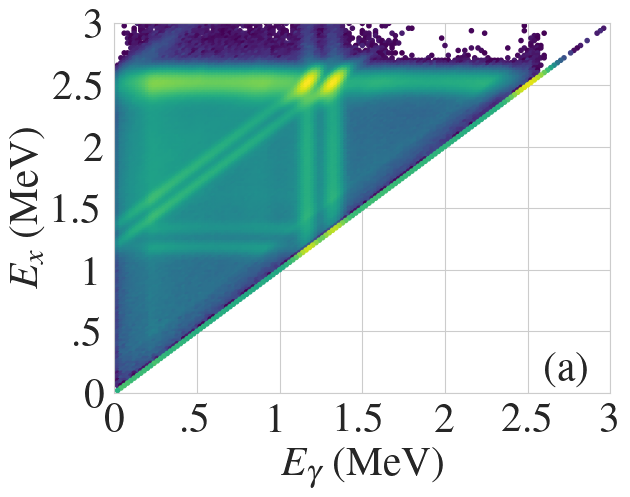}\hfill
\includegraphics[width=.3\textwidth]{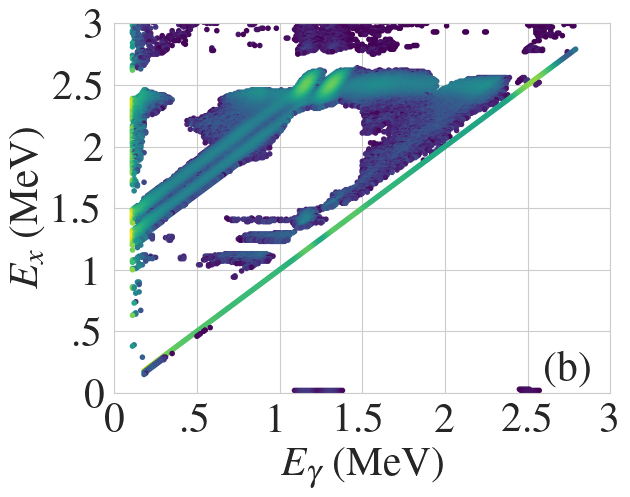}\hfill
\includegraphics[width=.3\textwidth]{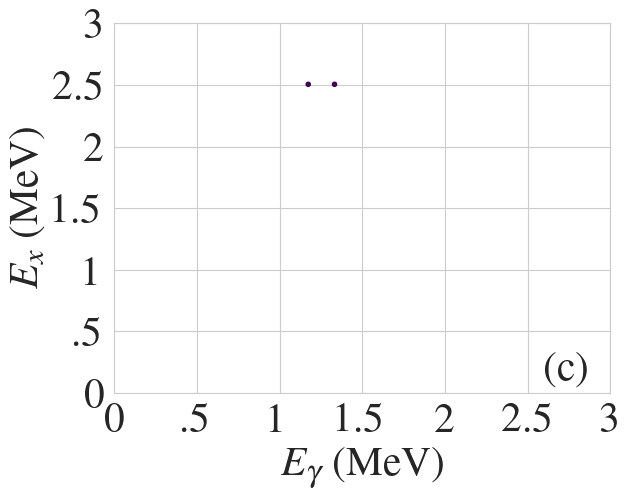}

\caption{Example spectra for the $\beta$-decay of $^{60}$Co. Subfigure (a) shows the decay as measured by SuN. Subfigure (b) shows the measured spectrum following traditional unfolding. Subfigure (c) shows the ideal measurement.}
\label{fig:unfolded}

\end{figure}

\cade{An ideal \rawmatrix~measurement of this decay would feature only two discrete points at locations corresponding to these two $\gamma$-ray energies on the $x$-axis and the single $2505$ keV excitation energy on the $y$-axis. Realistic measurements are complicated by factors such as imperfect detector resolution and $\gamma$-matter interactions during detection (such as Compton scattering and pair-production), and contain a continuous distribution of counts concentrated around the true energy locations. TAS measurements are further complicated by the fact that distortions in the measurement of $E_\gamma$ are correlated with distortions in the measurement of $E_x$. 
For accurate physical analyses, it is thus essential to ``unfold'' such data by removing the effects of detector response --- a problem that is simpler in theory than in practice.}  An ideal unfolding method would translate a dense \rawmatrix~matrix to a corresponding sparse \unfoldedmatrix~matrix that contains counts at locations corresponding to level scheme decays \cade{(\ref{fig:unfolded}).}

Methods such as those described in \cite{Guttormsen1996} have proven most successful in the past, and are readily available to the community via software packages like the Oslo Cyclotron Lab's Matrix Manipulation (MaMa) package \cite{Guttormsen_mama}. These unfolding procedures are widely used, including for the SuN detector, but are uniaxial and only partially eliminate correlational effects in the \rawmatrix~matrices. Figure~\ref{fig:unfolded} shows the shortcomings of this method; while there is improvement in reducing the prevalence of vertical and horizontal tails in the data, diagonal correlations are unaffected and stand in sharp contrast to the ideally unfolded matrix.

Alternative methods to combat these challenges are thus of great interest to the $\gamma$-ray spectroscopy community, and have been explored in the past. In particular,
automated unfolding shows promise in eliminating distortions invisible to traditional methods --- early attempts at machine learning-based analysis methods for one-dimensional $\gamma$-ray spectra data date back almost thirty years \cite{KoohiFayegh1993}. Recent advances in deep learning have begun to realize the feasibility of this style of approach on a broad scale \cite{Bailey2021,Gladen2020,Jhung2020,Kamuda2019,Kamuda2020,Medhat2012}.
We present a method of unfolding two-dimensional \rawmatrix~matrices via a conditional Generative Adversarial Network (cGAN). cGANs have been applied to a range of image translation and reconstruction problems in physics and biomedical imaging \cite{Fysikopoulos2021, Kim2020,List2019, Velasco2022,Shahnawaz2021}, which indicate potential for $\gamma$-ray spectroscopy applications.

\section{Machine Learning Methods}
\label{sec:ml}

\subsection{Unfolding \rawmatrix~Matrices As Image-to-Image Translation} 

Suppose we are given a training set of ordered pairs $\mathcal{T} = \{(i, i'): i \in I~\text{and}~i' \in I'\}$, where $I$ and $I'$ are sets of images from two domains. As a concrete example, one could consider pairs $(i, i')$ where the samples $i$ are drawn from daytime urban scenes, while the samples $i'$ are the same scenes after dark. The image-to-image translation problem seeks to fit a parameterized function $f: I \mapsto I'$ that converts an instance $i \in I$ into an instance $i' \in I'$ --- in our example, $f$ would take a daytime scene and transform it into a nighttime one. \citeauthor{Isola2016}'s landmark publication \cite{Isola2016} was one of the earliest works to demonstrate the capacity of deep neural networks to learn such functions $f$ from data. We cast the problem of unfolding \rawmatrix~matrices as an instance of image-to-image translation. Specifically, we create a dataset comprising pairs $(i, i')$ where each $i$ is a raw \rawmatrix~matrix and $i'$ is its corresponding unfolded matrix \unfoldedmatrix, which serves as the training set for our image-to-image translation model.

\subsection{Conditional GANs}
We train a conditional Generative Adversarial Network (cGAN) to solve our image-to-image translation problem. A cGAN typically comprises two neural networks engaged in an adversarial game: a \emph{generator} and a \emph{discriminator}. The generator's task is to create outputs that look realistic (i.e., that look like they may have been sampled from the target distribution), while the discriminator attempts to tell apart these ``fake'' samples from real ones drawn from the training data. Formally, the generator is a neural network $G: I \times Z  \mapsto I'$ which models the conditional distribution $\Pr(i' \mid i)$, where $i \in I$ is a raw matrix \rawmatrix~and $i' \in I'$ is an unfolded matrix \unfoldedmatrix. The additional input to $G$, $\mathbf{z} \in Z$, implements a trick that is widely used to endow neural networks with stochastic behavior: $\mathbf{z} = [z_0, z_1, \ldots, z_n]$ is a noise vector whose components $z_i$ are independently sampled from a standard distribution like $\mathcal{N}(0, 1)$. By computing $G(i, \mathbf{z})$ for a fixed $i$ and different values of $\mathbf{z}$, one can obtain multiple samples from $\Pr(i' \mid i)$. The discriminator is a separate neural network $D: I' \mapsto [0, 1]$. Given an unfolded matrix $i' \in I'$, $D$ outputs a score indicating the network's belief in whether $i'$ came from the training data or was generated by $G$. Higher values correspond to increased confidence that the input was a ``real'' sample drawn from the training set.

\subsection{The \pixtopix~Architecture}
We now describe \pixtopix~\cite{Isola2016, Tutorial}, the cGAN architecture that we use in this work. Our model development was guided by \cite{Tutorial}, a freely available tutorial on how to build cutting-edge Pix2Pix cGANs  provided by the developers of Tensorflow \cite{Tensorflow2015}. Our generator and discriminator utilize the same architectures and losses as presented within, with modifications made only to input sizes, discriminator patch resolution, and generator output activation (changed from tanh to sigmoid). A summary of the architectures is presented below.

The discriminator $D$ is inspired by the PatchGAN architecture first described by \citeauthor{Isola2016} \cite{Isola2016}. The key innovation in PatchGAN is to offer feedback to the generator at a more localized scale, by scoring \emph{patches} --- smaller regions of the input matrix ---  rather than the entire input. Further, the neural network operates on a pair of matrices $(i, i')$ as input, so that the patch scores evaluate the quality of the translation in different parts of the matrices.  Concretely, $D: I \times I' \mapsto [0, 1]^{p \times p}$, where $p$ is the patch resolution. We fit the parameters of the discriminator by minimizing the following loss function
\begin{equation}
    \mathcal{L}_D(D, G, \mathcal{T}) = -\frac{1}{n} \sum_{(i,i') \in \mathcal{T}} \sum_{\pi \in P} \log(D_{\pi}(i,i')) + \log(1 - D_{\pi}(i,G(i)))
\end{equation}
where $n$ is the number of examples in our training set $\mathcal{T}$, $P$ denotes the set of all patches and $D_{\pi}$ is the score assigned by the discriminator to patch $\pi$.

The generator $G$ is a modified implementation of the UNet autoencoder \cite{Ronneberger2015}.  UNet-style architectures have been shown to be both computationally efficient and effective at image translation problems \cite{Ronneberger2015}, and a similar variation was used in the original \pixtopix~system \cite{Isola2016}. Our generator deviates from the traditional cGAN formulation in one key way: since our mapping problem is completely deterministic (i.e., there is exactly one unfolded matrix $i'$ that corresponds to an input matrix $i$), we eliminate the noise vector $\mathbf{z}$ as an input to the generator, so that $G$ simply maps elements of $I$ to elements of $I'$.

The parameters of the generator are fit by minimizing a loss function $\mathcal{L}_G$ given by
\begin{equation}
\mathcal{L}_{G}(D,G,\mathcal{T})=\mathcal{L}_{adv}(D,G,\mathcal{T}) + \lambda \cdot \mathcal{L}_{L1}(G,\mathcal{T})
\label{lab:gen-loss}
\end{equation}
where
\begin{equation}
\mathcal{L}_{adv}(D,G,\mathcal{T}) = -\frac{1}{n}\sum_{(i,i') \in \mathcal{T}} \sum_{\pi \in P} \log(D_{\pi}(i, G(i)))
\end{equation}
and
\begin{equation}
\mathcal{L}_{L1}(G,\mathcal{T}) = \frac{1}{n} \sum_{(i,i') \in \mathcal{T}}|i'-G(i)|.
\end{equation}
The generator loss $\mathcal{L}_G$ comprises two components: an adversarial term ($\mathcal{L}_{adv}$) and an L1-norm term ($\mathcal{L}_{L1}$). Minimizing the former corresponds to the generator's samples ``fooling'' the discriminator into believing that they are genuine unfolded matrices drawn from the training data. Minimizing the latter ensures that the outputs of the generator are objectively close to the ground truth unfolded matrices in the dataset, under a traditional distance metric. We also considered other candidates, such as the $L_2$-norm, in this second term, but we found that the $L_1$ loss consistently outperformed the others.

$\mathcal{L}_{L1}$ is scaled by a factor $\lambda$ to the same magnitude of $\mathcal{L}_{adv}$. Because the sparsity of our matrices leads to low values of $\mathcal{L}_{L1}$, we found a value of $300$ to be most effective for $\lambda$, approximately a factor of three times higher than \cite{Tutorial}.

This value, as well as those for other hyperparamters, are reported in \ref{table:hyperparameters}.
 
We use the Adam \cite{Kingma2017} algorithm for training both the discriminator and the generator. Adam is a member of the stochastic gradient descent family that utilizes parameter-specific learning rates based on the magnitude of recently calculated gradients. It is a standard choice for training deep learning models due to its stability and speed of convergence under a wide range of conditions \cite{Ruder2016}.

\begin{table}[tb]
\centering
\begin{tabular}{||c|c||}
    \hline
    Epochs of training & $125$  \\
    \hline
    Batch size & $16$\\
    \hline
    Generator learning rate & $\num{2e-6}$\\
    \hline
    Discriminator learning rate & $\num{2e-6}$\\
    \hline
    Generator $\beta_{1}$ & $0.5$\\
    \hline
    Generator $\beta_{2}$ & $0.999$\\
    \hline
    Discriminator $\beta_{1}$ & $0.5$\\
    \hline
    Discriminator $\beta_{2}$ & $0.999$\\
    \hline
    Generator $\epsilon$ & $\num{1E-7}$\\
    \hline
    Discriminator $\epsilon$ & $\num{1e-7}$\\
    \hline
    $L_1$ loss scaling factor $\lambda$ & $300$\\
    \hline
    Patch resolution $p$ & $62$ \\
    \hline
\end{tabular}
\caption{The values of the hyperparameters used in our final model. The exponential decay rate for the first moment in the Adam optimizer is given by $\beta_1$, while $\beta_2$ is the exponential decay rate of the second moment. The term $\epsilon$ is a constant used to ensure numerical stability in Adam.}
\label{table:hyperparameters}
\end{table}


\section{Data Preparation}

\subsection{Data Simulation and Processing}
We used the GEANT4 \cite{Agostinelli2003} simulation package to simulate our training data. 
Each training spectrum contains a decay comprising either a single $\gamma$-ray or two $\gamma$-rays emitted in sequence, mimicking simple isotopic decay cascades. We choose a binning of 500 across a 0-10 MeV range along both axes. This binning is comparable to the resolution of the SuN detector \cite{Simon2013} and similar to binnings used in \cite{Spyrou2014, Liddick2016, Spyrou2017, Liddick2019}. 

We simulated a total of $9950$ single $\gamma$-ray decay spectra for each integer energy value within the $50$-$10000$ keV range. We further produced an additional $9451$ training spectra containing cascades of two $\gamma$-rays with randomly-generated energies, for a total of $19401$ training examples. The corresponding target spectra were generated as $500 \times 500$ arrays with pixel values of one denoting the location of the ground-truth $\gamma$-ray energies and zeroes in all other bins. The training and target spectra were padded with $12$ empty bins on their top and right edges to bring their size to $512 \times 512$, which allows for repeated downsampling through the UNet architecture's $2 \times 2$ pooling layers without numerical rounding issues.

\subsection{Train-Test Split and Standardization}

We followed standard machine learning methodology in estimating the generalization error of our trained models, by dividing our data into disjoint training and test sets. We used an $80$-$20$ split that yielded a total of $15520$ training spectra and $3881$ testing spectra. As typical for GANs, we trained our models until all losses converged, with particular attention paid to the minimization of $\mathcal{L}_{L1}$. The testing dataset was withheld and presented to the model post-training as a method of evaluating both its quality of output and generalization capabilities. A set of experimentally measured spectra were used as a final, definitive benchmark (see section \ref{sec:res}).

\begin{figure}[ht]
\captionsetup[subfigure]{labelformat=empty}
\begin{subfigure}{.5\textwidth}
  \centering
  
  \includegraphics[width=1\linewidth]{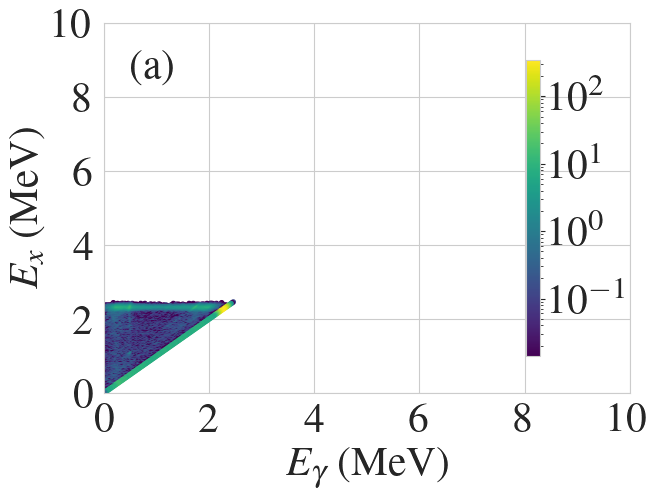}  
  \label{fig:1gamma}
  \caption{}
\end{subfigure}
\begin{subfigure}{.5\textwidth}
  \centering
  \includegraphics[width=1\linewidth]{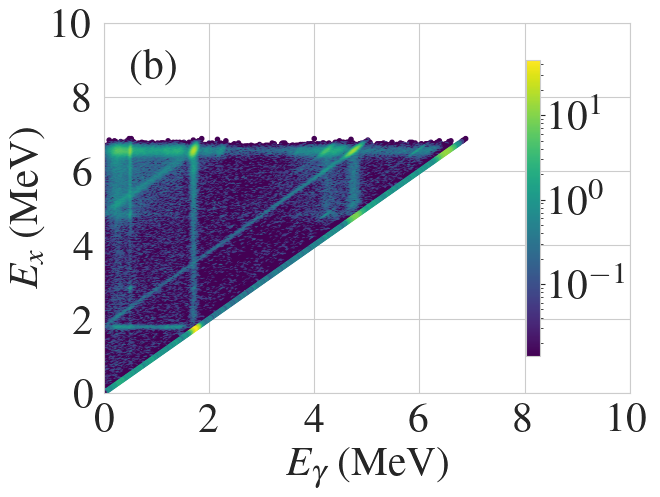}  
  \label{fig:2Gamma}
  \caption{}
\end{subfigure}

\vspace{-13.00mm}

\begin{subfigure}{.5\textwidth}
  \centering
 
  \includegraphics[width=1\linewidth]{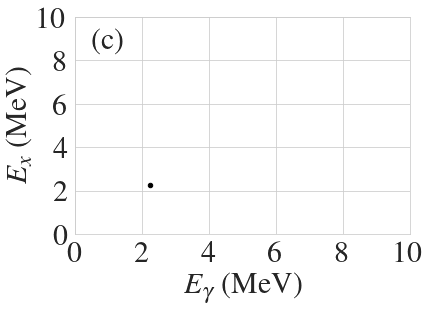}  
  \caption{}
  \label{fig:1gamma label}
\end{subfigure}
\begin{subfigure}{.5\textwidth}
  \centering
  
  \includegraphics[width=1\linewidth]{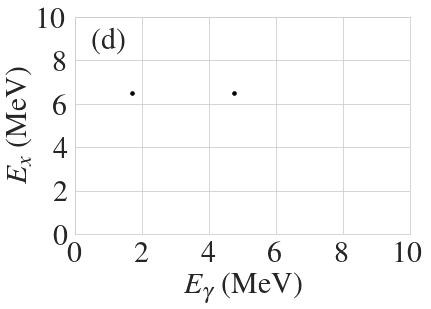}  
  \caption{}
  \label{fig:2gamma label}
\end{subfigure}

\vspace{-5.00mm}

\caption{Example training spectra and targets. Subfigures (a) and (b) respectively show single-$\gamma$ and double-$\gamma$ \rawmatrix~matrices from the standardized training set. Subfigures (c) and (d) show the the corresponding \unfoldedmatrix~matrices to (a) and (b).}
\label{fig:Train Data}
\end{figure}

The train and test sets were standardized to have a mean of zero and unit variance to account for the differences in detection efficiency across SuN's energy range. Example post-standardization training spectra and their corresponding labels are shown in Figure \ref{fig:Train Data}.

\section{Evaluation and Results}
\subsection{Evaluation Methodology}

To evaluate the quality of our machine learning-based unfolding procedure, we need a method to compare the spectra output by our model to the target spectra. \cade{This task is complicated by the nature of the problem.} Standard distance metrics, like computing the $L_1$ or $L_2$ norm between the predicted and target spectra, conflate different sources of error --- for example, a prediction that places a single \cade{nonzero matrix entry} one bin away from the correct location could be deemed just as bad as one that places the prediction several MeV away, or one that smears out its predictions over a larger region of energy-space. \cade{This makes errors very challenging to interpret.} A different option might be to use $\mathcal{L}_{G}$ as our performance metric. This choice has the advantage that it measures the quantity that our generator is directly optimizing. However, this metric \cade{also} suffers from the drawback that its value is not directly interpretable from a physics standpoint. To circumvent these issues, we use a two-step evaluation strategy: we first consolidate the output from the model using a clustering algorithm, and \cade{then compare the locations of the cluster centroids to the ground-truth $\gamma$-ray locations.} We now describe the details of this procedure.

\subsubsection{$k$-means Clustering}
The $k$-means clustering problem seeks to group a set of $n$ sample observations $\{\bm{x_1}, \bm{x_2}, \ldots, \bm{x_n}\}$ into $k$ disjoint clusters $\{C_1, C_2, \ldots, C_k\}$ (where $k \leq n$) in a manner that minimizes the variance among the members of each cluster. Each cluster $C_j$ is described by the mean of the points assigned to it, denoted by $\bm{\mu_j}$. Formally, the aim is to solve the following optimization problem:
$$ \argmin_{\bm{\mu_j}} \sum_{i=1}^{n} \sum_{j=1}^{k} \mathbbm{1}[x_i \in C_j] \cdot || \bm{x_i} - \bm{\mu_j} ||^2 $$
While solving this problem optimally is \textsf{NP}-hard, heuristic approaches such as Lloyd's algorithm \cite{lloyd}, which iteratively determines the centroids of the $k$ clusters, are often effective in practice. 

\subsubsection{Clustering Predicted Spectra}
We perform two post-processing steps on a matrix output by our model to enable a meaningful comparison to the target spectrum. First, we filter out noise by rounding down to $0$ all matrix entries that are less than a cutoff threshold of $0.2$.
We then apply $k$-means clustering to the coordinates of the non-zero entries in this filtered matrix, with $k$ set to either $1$ or $2$ depending on the number of $\gamma$-rays in the true spectrum. We use a weighted variant of Lloyd's algorithm in performing this clustering, using the entries in the matrix as the weights. This approach treats a matrix entry at location $(E_x, E_\gamma)$ as a measure of the model's confidence in  $(E_x, E_\gamma)$ being the location of a $\gamma$-ray. We use the implementation of Lloyd's algorithm provided by the \texttt{scikit-learn} Python library \cite{Pedregosa2011}.

\subsubsection{Percentage Error}
The centroid(s) returned by the $k$-means analysis are treated as the model's effective prediction. Percent errors in $E_x$ and $E_{\gamma}$ are then calculated based on this prediction and the ground-truth $\gamma$-ray location, which allows for direct comparison between the model's results and SuN's resolution capabilities \cite{Simon2013}. As a Sodium Iodide scintillator, SuN's resolution is inherently limited, generally to within $5$--$7\%$ of the energy of the measured $\gamma$-ray. The resulting width of SuN's energy peaks affects both the bin distribution of counts in the input spectra and the accuracy limitations on model predictions and percent error metrics provide a reasonable method for determining if a given $\gamma$-ray prediction falls within these limitations.


\subsection{Results}
\label{sec:res}

\begin{figure}[ht]
\captionsetup[subfigure]{labelformat=empty}

\begin{subfigure}{.5\textwidth}
  \centering
  \hspace*{3.2cm}\includegraphics[width = 0.6 \linewidth]{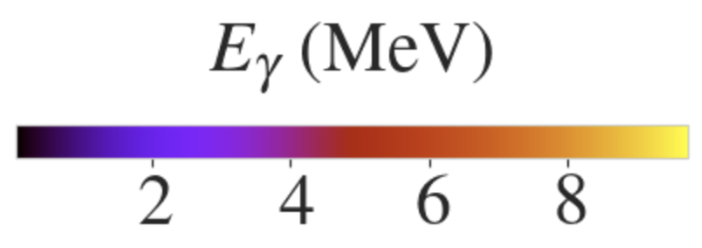}\hspace*{-3.2cm} 
  \caption{}
  \label{fig:Singles_Unfolded}
\end{subfigure}

\hspace{-3.00mm}
\begin{subfigure}{.5\textwidth}
  \centering
  \includegraphics[width=1.025\linewidth]{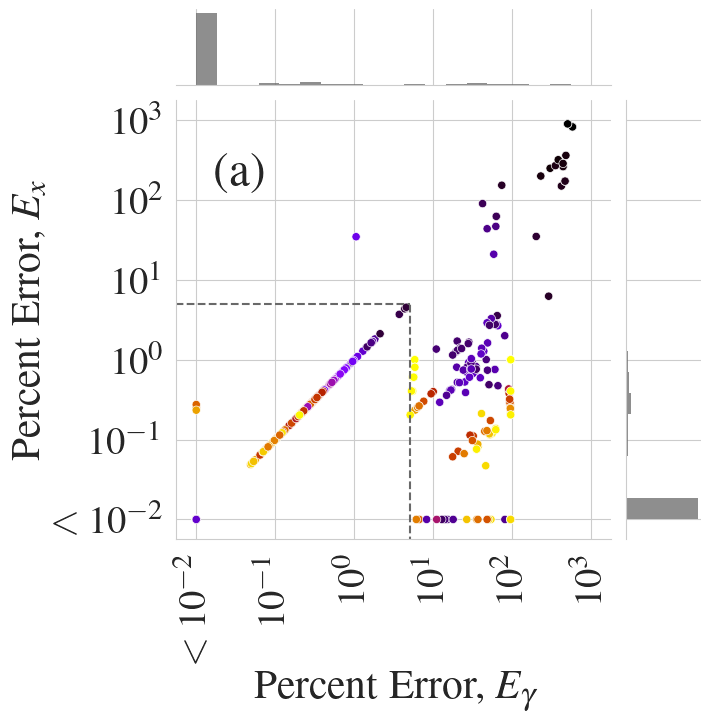}  
  \caption{}
  \label{fig:Singles_Unfolded}
\end{subfigure}
\hspace{1mm}
\begin{subfigure}{.5\textwidth}
  \centering

  \raisebox{-4.7cm}{\includegraphics[width=1.025\linewidth]{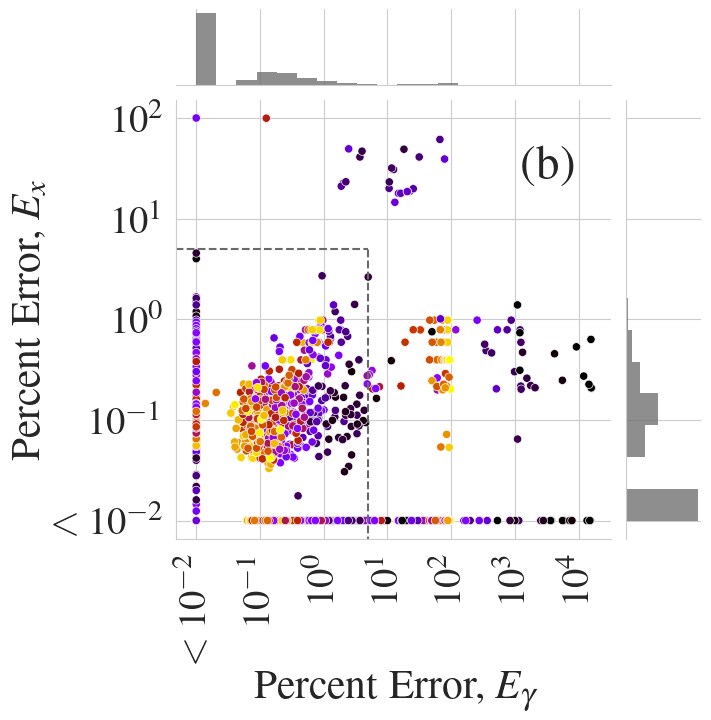}}  
  \caption{}
  \label{fig:Doubles_Unfolded}
\end{subfigure}

\begin{subfigure}{.5\textwidth}
  \centering
 
  \includegraphics[width=1\linewidth]{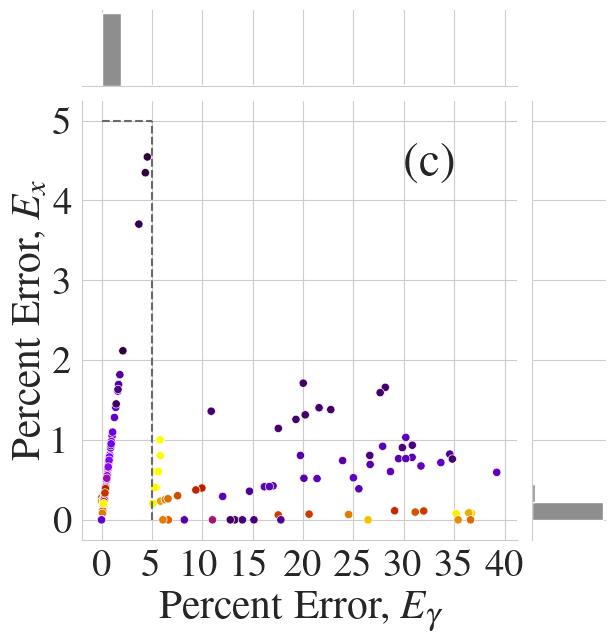}  
  \caption{}
  \label{fig:Singles_Unfolded_Zoom}
\end{subfigure}
\begin{subfigure}{.5\textwidth}
  \centering
 
  \includegraphics[width=1\linewidth]{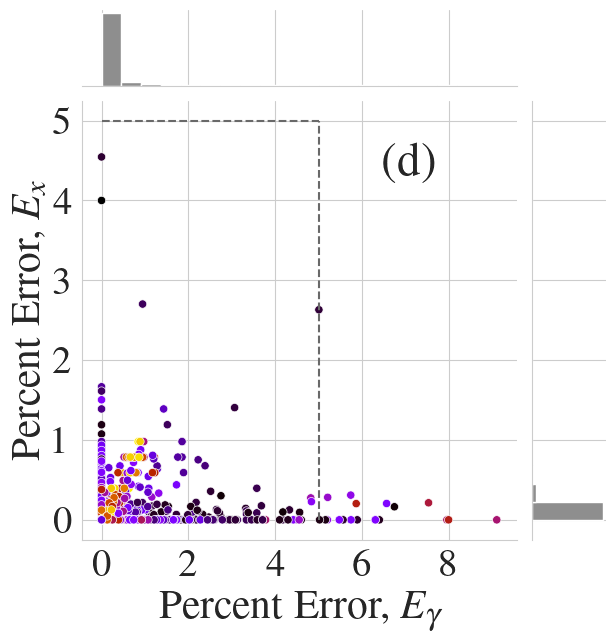}  
  \caption{}
  \label{fig:Doubles_vs_Energy_Zoom}
\end{subfigure}

\caption{Evaluation results for the testing dataset of simulated spectra. Datapoints are colored by their ground-truth $\gamma$-ray energy, Subfigures (a) and (b) show results for single and double $\gamma$-ray test spectra after machine-learning unfolding. Subfigures (c) and (d) show results the same results, but with the axes cut to show two standard deviations worth of data. \cade{Grey histograms show the density of entries along each axis. Dashed lines show lower resolution limits for SuN.}}
\label{fig:Testing Results}
\end{figure}

Table~\ref{table:hyperparameters} presents the hyperparameter settings that resulted in the best performing model in our experiments. Code and data to train and evaluate these models, as well as a pre-trained model, are available at \cite{GitHub}. \cade{Figure \ref{fig:Testing Results} shows} the results of the $k$-means clustering analysis for the cGAN model's testing dataset predictions, with percent error in $E_{x}$ and $E_{\gamma}$ evaluated for each $\gamma$-ray present within the testing spectra. \cade{Histograms showing the relative density of examples are provided on the axes of each plot to show how predictions are distributed, which would not otherwise be evident.} \cade{Across all testing spectra, 93.2$\%$ of $\gamma$-ray predictions fell within 5$\%$ of the ground truth in both $E_x$ and $E_{\gamma}$ after unfolding. This demonstrates the cGAN's effectiveness in removing distortions from detector response effects while retaining accurate $\gamma$-ray signals for the  majority of cases.}

\cade{The outliers in Figure \ref{fig:Testing Results} are quite noticeable, and further discussion of this point is beneficial.
Manual investigation has shown that common failures include events like the incorrect filtering of a true $\gamma$-ray peak or retention of erroneous signals like pair-production $\gamma$-rays. In these cases, the clustering algorithm is comparing a different $\gamma$-ray prediction to the ground truth, generally resulting in a high percent error that loses its meaning.
Despite this, the clustering  method is a useful evaluation tool given its interpretive benefits for the preponderate successful predictions.}



\begin{figure}[ht]
\captionsetup[subfigure]{labelformat=empty}
\begin{subfigure}{.5\textwidth}
  \centering
  
  \includegraphics[width=\linewidth]{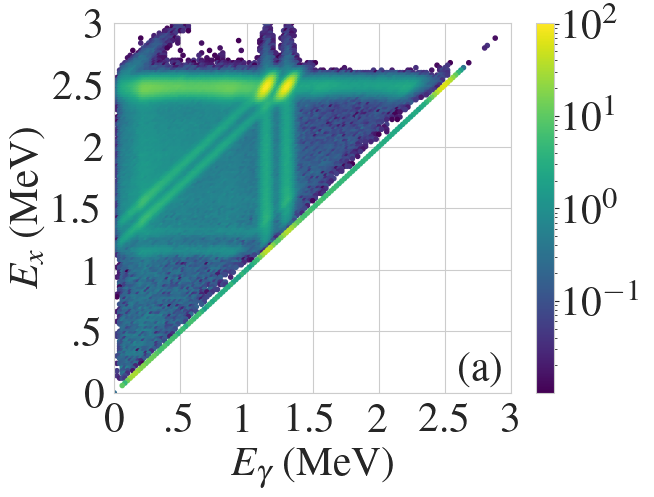}  
  \caption{}
  \label{fig:Co raw}
  
\end{subfigure}
\begin{subfigure}{.5\textwidth}
  \centering
  
  \includegraphics[width=\linewidth]{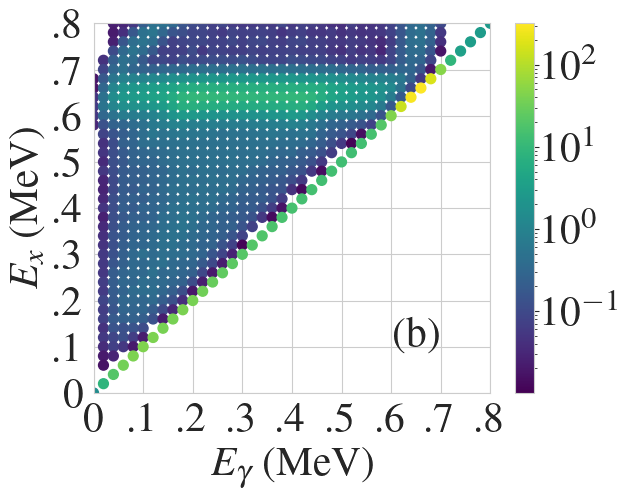}  
  \caption{}
  \label{fig:Cs Raw}
\end{subfigure}

\vspace{-6.00mm}

\begin{subfigure}{.5\textwidth}
  \centering
 
  \includegraphics[width=1\linewidth]{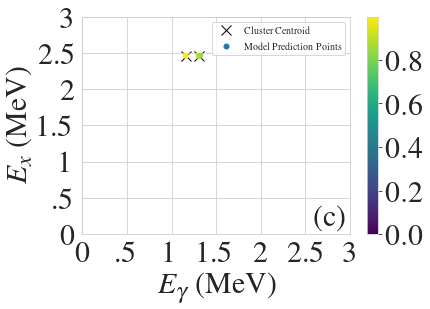}  
  \caption{}
  \label{fig:Co Prediction}
\end{subfigure}
\begin{subfigure}{.5\textwidth}
  \centering
  
  \includegraphics[width=1\linewidth]{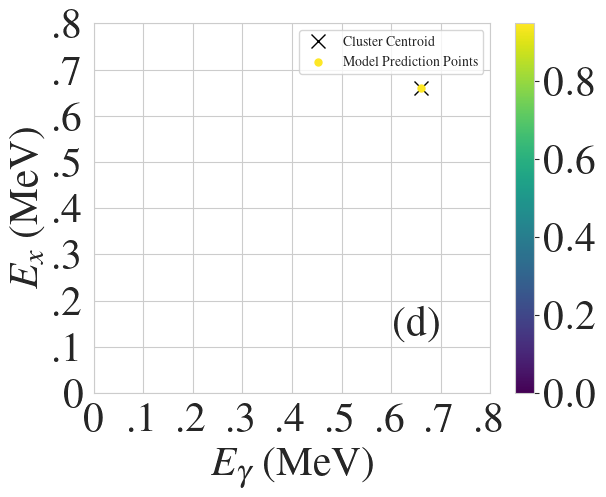}  
  \caption{}
  \label{fig:Cs Prediction}
\end{subfigure}

\begin{subfigure}{.5\textwidth}
  \centering
  
  \includegraphics[width=1\linewidth]{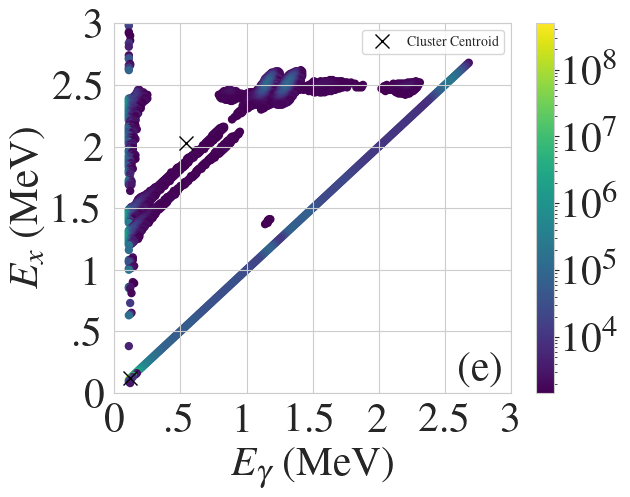}  
  \caption{}
  \label{fig:Co Unfolded}
\end{subfigure}
\begin{subfigure}{.5\textwidth}
  \centering
  
  \includegraphics[width=1\linewidth]{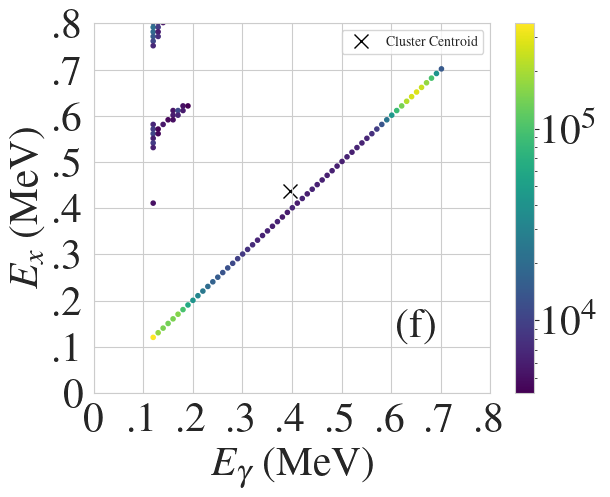}  
  \caption{}
  \label{fig:Cs Unfolded}
\end{subfigure}
\caption{Standardized input and predicted output spectra for experimental data on $^{60}$Co and $^{137}$Cs decays. Subfigures (a) and (b) show standardized \rawmatrix~input matrices for $^{60}$Co and $^{137}$Cs respectively. Additional counts at energies higher than the true decay result from SuN's detection of more than one event within a single timing window. Subfigures (c) and (d) show corresponding model predictions with $k$-means centroids overlaid (cutoff threshold of .04). Subfigures (e) and (f) show a comparison to the traditional unfolding method, also with overlaid $k$-means clusters (the same cutoff threshold as (c) and (d), scaled by the number of counts in $\gamma$-ray peaks). Here, subfigures (a) and (b) are zoomed in to highlight salient features, but the full spectra extend to the same 10MeV limits as the training dataset.}
\label{fig:Exp Results}
\end{figure}

\cade{As another test, }we also investigated model performance on data sourced from actual experiments. We evaluate the model using \rawmatrix~matrices for the decays of common $\gamma$-ray sources $^{60}$Co and $^{137}$Cs placed at the center of the SuN detector. $^{137}$Cs predominantly $\beta$-decays to an isomeric state in $^{137}$Ba that then decays via emission of a single $\gamma$-ray at $662$ keV. The decay scheme of $^{60}$Co is as discussed in \ref{Intro}. These results are shown in Figure \ref{fig:Exp Results}. The $662$ keV $\gamma$-ray of the $^{137}$Cs decay is predicted within bin accuracy. The $1173$ keV $\gamma$-ray of the $^{60}$Co decay is predicted within bin accuracy for $E_{\gamma}$, and with an error of $1.80 \%$ for $E_x$. The $1332$ keV $\gamma$-ray is predicted with $1.70 \%$ and $1.80 \%$ errors for $E_{\gamma}$ and $E_x$, respectively.

While it is impractical to apply the traditional unfolding method described in \ref{Intro} to the thousands of simulated spectra in the testing dataset, application to the experimental source spectra illuminates the benefits of the machine-learning based approach. The traditional unfolding method suffers from an accumulation of a large number of counts along the edges of the matrix - particularly at low values of excitation and $\gamma$-ray energy. An artificial ``peak'' resulting from this effect, located between 100-200 keV along each axis, features in Figures~\ref{fig:Co Unfolded} and \ref{fig:Cs Unfolded}. This method's failure to remove diagonal tail artifacts is also evident. The cGAN-unfolded spectra, shown in Figures~\ref{fig:Co Prediction} and \ref{fig:Cs Prediction}, do not exhibit these problems and still retain energy information on a level of accuracy comparable to the detector resolution.

\section{Conclusions and Future Work}

We have demonstrated the effectiveness of conditional GANs in constraining simple nuclear level schemes from ($E_{x}$,$E_{\gamma}$) matrices. Our trained cGAN model showed prediction capabilities comparable to the energy resolution of the SuN detector for over $93\%$  of $\gamma$-rays in a statistically independent testing dataset. Additionally, the model was shown to accurately characterize the level schemes of common sources $^{60}$Co and $^{137}$Cs from experimentally measured \rawmatrix~matrices.

As it stands, this work is a promising proof of concept for the use of conditional GANs in total absorption spectroscopy analysis and future efforts are planned in several areas. In order to be useful for the analysis of unstable nuclei far from the valley of stability, the model must be trained on increasingly complex decay schemes with more $\gamma$-rays of varying intensities. Other improvements to the $k$-means evaluation process, like the implementation of a convolutional neural network to determine the number of $\gamma$-rays present in an output spectrum, will allow for completely automated analysis of unknown level schemes. Taken together, these future plans will expand the model's applicability, building upon the promise of the work presented here.

\section{Acknowledgements}
We would like to thank Andrew Hoyle for his discussions about our architecture choice for this project. The work was supported by the National Science Foundation under grants
PHY 2012865, 
PHY 1913554,  
PHY 1430152, 
PHY 1613188. 
 This work was supported by the US Department of Energy (DOE) National Nuclear Security Administration Grant No DOE-DE-NA0003906 and the DOE Office of Science under Grant No. DE-SC0020451. This material is based upon work supported by the Department of Energy/National Nuclear Security Administration through the Nuclear Science and Security Consortium under Award No. DE-NA0003180.
 This work was supported by computational resources provided by the Institute for Cyber-Enabled Research at Michigan State University.





\clearpage{}

\bibliographystyle{elsarticle-num-names}
\typeout{}
\bibliography{els_classification}







\end{document}